\documentclass[twocolumn,english,showpacs,showkeys,reprint]{revtex4}
\usepackage[T1]{fontenc}
\usepackage[latin9]{inputenc}
\usepackage{color}
\usepackage{textcomp}
\usepackage{amsmath}
\usepackage{graphicx}

\makeatletter
\@ifundefined{textcolor}{}
{%
 \definecolor{BLACK}{gray}{0}
 \definecolor{WHITE}{gray}{1}
 \definecolor{RED}{rgb}{1,0,0}
 \definecolor{GREEN}{rgb}{0,1,0}
 \definecolor{BLUE}{rgb}{0,0,1}
 \definecolor{CYAN}{cmyk}{1,0,0,0}
 \definecolor{MAGENTA}{cmyk}{0,1,0,0}
 \definecolor{YELLOW}{cmyk}{0,0,1,0}
}

\usepackage{babel}

\usepackage{babel}

\usepackage{babel}

\usepackage{babel}

\usepackage{babel}

\makeatother

\usepackage{babel}
\begin{document}

\title{\textcolor{black}{Voltage-induced strain control of the magnetic
anisotropy in a Ni thin film on flexible substrate}}

\author{F.Zighem\footnote{zighem@univ-paris13.fr}, D. Faurie, S. Mercone, M. Belmeguenai and H. Haddadi}

\affiliation{LSPM-CNRS, Université Paris XIII, Villetaneuse, France}

\begin{abstract}
Voltage-induced magnetic anisotropy has been quantitatively studied
in polycrystalline Ni thin film deposited on flexible substrate using
microstrip ferromagnetic resonance. This anisotropy is induced by
a piezoelectric actuator on which the film/substrate system was glued.
In our work, the control of the anisotropy through the applied elastic
strains is facilitated by the compliant elastic behavior of the substrate.
The in-plane strains in the film induced by the piezoelectric actuation
have been measured by the digital image correlation technique. Non-linear
variation of the resonance field as function of the applied voltage
is found and well reproduced by taking into account the non linear
and hysteretic variations of the induced in-plane strains as function
of the applied voltage. Moreover, we show that initial uniaxial anisotropy
attributed to compliant substrate curvature is fully compensated by
the voltage induced anisotropy.
\end{abstract}


\maketitle

\section{Introduction}

Prospect of controlling local magnetization using electric fields,
for low power and ultra fast new electronics has resulted in an
amount of new research domains mainly focused on artificial
engineered materials. Two decades ago, the condensed matter
community began studying the interactions between a spin polarized
current \cite{Slonczewski1996,Kiselev2003} and magnetic domain walls
in magnetic thin films. Percolation and domain wall excitation
(precessional and/or translational) by spin angular momentum
transfer from the current to the magnetic system pushed the study of
new magnetic materials and advanced lithographed magnetic devices
\cite{Parkin2007,Mercone2002,Mercone2003}. Since then, new magnetic
configurations have been engineered by properly shaping magnetic
nanostructures (for instance: U-shaped nanowires, focused ion beam
nanobridges, giant magnetoresistance spin valve nanowires systems,
magnetic racetrack memory devices). Unfortunately many experimental
studies have indicated extremely low current-driven domain wall
velocity and also non-adiabatic contributions to
current-magnetization interaction, which introduce intrinsic
threshold current for domain wall motion. Clearly, either the
maximum possible speed of domain wall based driven recording and
storage systems or power dissipation (directly linked to the current
threshold) are not suitable for ultra-fast future electronics.

In order to overcome these important factors limiting the future miniaturization
of integrated recording circuits based on simple magnetic nanostructured
material, latest approaches start to involve intrinsic multiferroic
and magnetoelectric artificial materials. Natural multiferroics have
quite immediately shown a too weak magnetic order at room temperature
making them useless for applications \cite{Cheong2007,Skumryev2011,Spaldin2005}.
Using artificial nanostructures combining the two different orders
(magnetic and electric one) in a unique system seems to be the only
promising route. The easiest artificial achitecture for a voltage
control of the magnetization in those kinds of materials, appears
to be a piezoelectric/magnetostrictive bilayers presenting a good
strain-mediated coupling at the interface \cite{Eerenstein2007,Lecoeur2011,Vaz2010}.
In order to achieve the highest magnetoelectric coupling for an efficient
voltage manipulation of the magnetization, the magnetic device has
to be encapsulated inside a piezoelectric environment which lead to
two main difficulties. First, deposition of piezoelectric material
generally needs high temperature and the most of the time also post
annealing treatment under severe condition of oxidation. This latter
could degrade the good magnetic properties of the magnetic layer.
Second, the clamping effect due to the substrate, limits the stress
applicable to the magnetization by the piezoelectric environment reducing
perspective on applications. Indeed, in this kind of bilayers (piezoelectric/magnetostrictive),
a significant strained induced magnetoelectric coupling is obtained
only in the presence of non negligible in-plane strains. In order
to avoid both latters issues, inverse magnetoelectric effect has been
studied by depositing ferromagnetic thin films on a piezoelectric
substrate \cite{Liu2010,Lou2009,Wang2012,Li2012,Pettiford2008,Brandlmaier2009,Brandlmaier2011,Brandlmaier2012}
or by directly gluing (\textit{via} an epoxy glue) a (thin film/ rigid
substrate) magnetic system onto a piezoelectric actuator \cite{Brandlmaier 2008,Chen2011,Chen2009,Lou2008}.

Noticeably, Brandlmaier et \textit{al}.\cite{Brandlmaier 2008} have
shown that film strains are much smaller (50\%) than the expected
values in case of a complete strain transmission through the cement
and the rigid substrate.

In this paper, voltage induced in-plane magnetic anisotropy has been
quantitatively studied in a magnetostrictive film deposited on a flexible
substrate and glued onto a piezoelectric actuator. For this study,
Nickel has been chosen due to its well-known negative effective magnetostriction
coefficient at saturation even in a polycrystalline film with no preferred
orientations. Digital Image Correlation (DIC) and Micro-Strip FerroMagnetic
Resonance (MS-FMR) techniques have been employed to i) quantitatively
measure the in-plane strains transmitted to the ferromagnetic film
responsible of the inverse magnetoelectric effect occurring in the
fabricated biferroic structure and to ii) experience the voltage-driven
changes of the in-plane magnetic anisotropy.

\section{Experimental details}

\begin{figure}
\includegraphics[bb=30bp 180bp 820bp 540bp,clip,width=8.5cm]{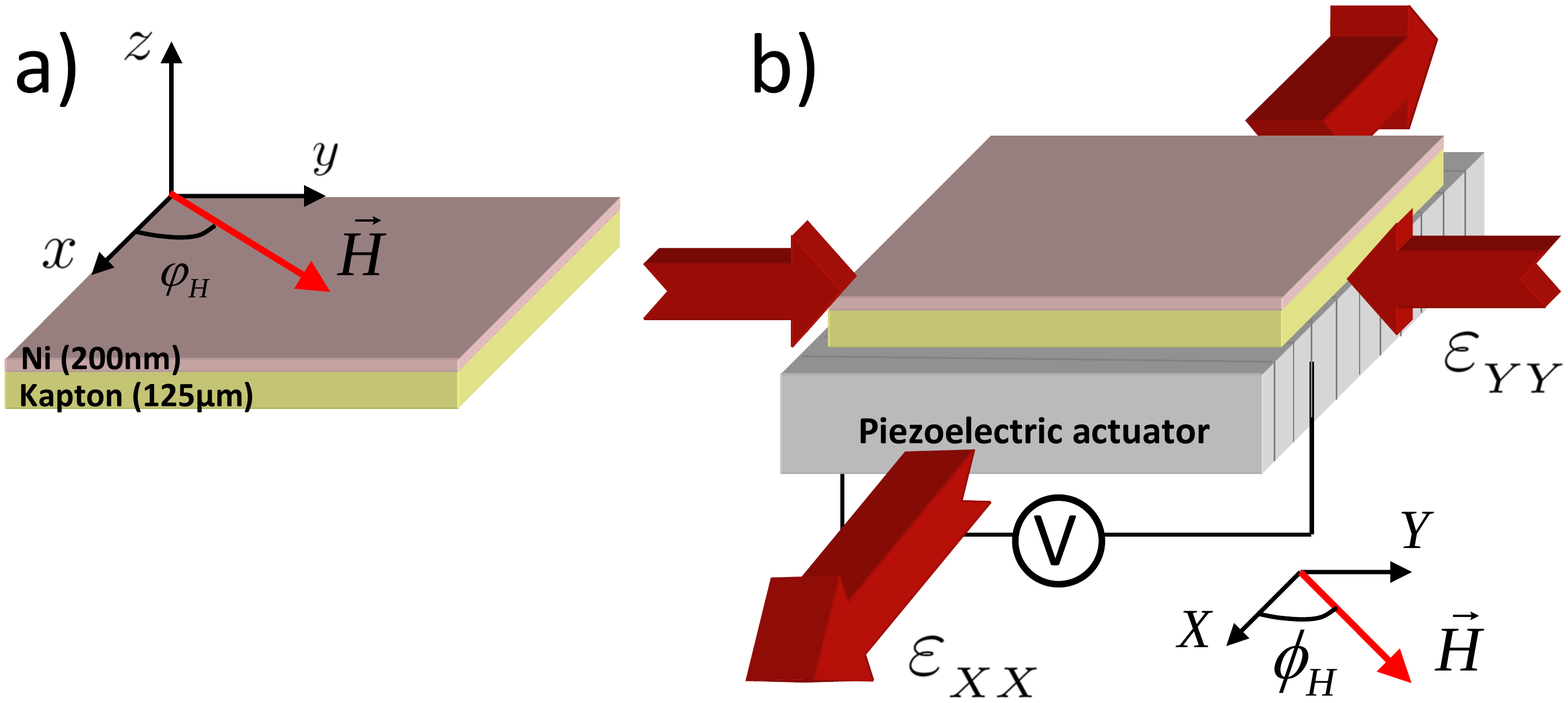}

\caption{a) Sketch of the 200 nm thick Ni thin film deposited onto a
Kapton\textregistered{} substrate (125 $\text{\textmu}$m). b)
Schematic diagram depicting the applied voltage within the
Piezoelectric/Kapton\textregistered{}/Ni heterostructure. Note that
two different coordinate systems are used; $(x,y,z)$ and $(X,Y,Z)$
respectively related to the Ni thin film and to the piezolelectric
actuator. Thereby, $\varphi_{H}$ is the angle between the applied
magnetic field $\vec{H}$ and $x$ direction and $\phi_{H}$ is the
angle between $\vec{H}$ and $X$ direction. }
\end{figure}

A 200 nm-thick Nickel film was deposited on a flexible (Kapton\textregistered{}
\cite{Kapton}) substrate (125 $\text{\textmu}$m) by radio frequency
sputtering at room temperature. The Ni film was found to be polycristalline
with no strong preferred orientations by x-ray diffraction. After
deposition, the film/substrate system was glued onto a piezoelectric
actuator. A flexible compliant substrate (Young's modulus $E=4$ GPa)
has been chosen in order to avoid clamping effects which can occur
with rigid substrates (for example, $E=180$ GPa for Silicon) and
thus to have maximum strain transmission in between the piezoelectric
actuator and the film \cite{Brandlmaier 2008,Chen2011,Lou2008}. Figure
1 shows a sketch of the Ni thin film and of the structure allowing
voltage-control of the in-plane magnetic anisotropy. Two different
axes systems are defined. The fi{}rst one $(x,y,z)$ is attached to
the Ni film while the second one $(X,Y,Z)$ describes the piezoelectric
actuator geometry. $\varphi_{H}$ (resp. $\phi_{H}$) is the angle
between $x$ (resp. $X$) axis and the applied magnetic field ($\vec{H}$);
the magnetization ($\vec{M}$) is defined by its polar angles ($\theta$
and $\varphi$) as usual \cite{belmeguenai2009}.

All the measurements presented here were performed at room temperature
using both microstrip ferromagnetic resonance (MS-FMR) and a digital
image correlation (DIC) technique. MS-FMR was employed to probe the
influence of the applied voltage on the magnetic properties \textit{via}
the resonance field of the uniform precession mode. The DIC was used
to quantitatively measure the in-plane strains ($\varepsilon_{XX}$
and $\varepsilon_{YY}$) induced by the piezoelectric actuator at
the surface of the ferromagnetic film.

MS-FMR experiments allowed the resonance field $H_{res}$ probing
by sweeping the applied magnetic field in presence of a fixed pumping
Radio Frequency (RF) fi{}eld $\vec{h}_{rf}$. In this kind of RF-resonance
technique, a weak modulation of the applied field ($\sim2-5$ Oe at
1480 Hz) is performed in order to optimize the signal to noise ration
for a better lock-in detection of the signal. Thus, this technique
gives access to the fi{}rst field derivative of the RF absorption
as function of the applied fi{}eld. A more detailed presentation can
be found elsewhere \cite{belmeguenai2013}.

DIC is a technique which allows displacement and strain fi{}elds measurements
on an object surface by capturing images of the object surface at
different states \cite{Haddadi2012,Djaziri2011}. One state is recorded
before applying voltage, i.e. the reference image, and the other states
are subsequent images of the deformed object. DIC generally uses random
patterns of gray levels of the sample surface to measure the displacement
\textit{via} a correlation of a pair of digital images. In this study,
the film surface was spray-painted with a speckle pattern to generate
a contrast in the uniform specimen face. Hence, from the measured
strain fields we can estimate the mean in-plane strains $\varepsilon_{XX}$
and $\varepsilon_{YY},$ that cannot be straightforwardly attained
with other techniques such as x-ray diffraction \cite{Djaziri2012}.

\section{Results and Discussion}

In the first paragraph of this section we present the magnetic parameters
of the Ni film at zero applied voltage. The second part is devoted
to a quantitative measure of the in-plane strains ($\varepsilon_{XX}$
and $\varepsilon_{YY}$) at the surface of the Ni film using the DIC
technique. In the third we present the analytical model allowing the
conversion of the applied voltage into strains, which we developped
to analyze the influence of the applied voltage on the magnetic properties
of the film observed fy MS-FMR.

\subsection{Magnetic parameters at zero applied voltage}

MS-FMR experiments in absence of applied voltage were performed in
order to determine the magnetic parameters of the Ni thin film. Indeed,
at zero voltage, the frequency of the uniform mode is expected to
only depend on the gyromagnetic factor $\gamma$, on the magnetization
at saturation $M_{S}$ and on the possible presence of magnetic anisotropies
\cite{Zighem2010}. Figure 2a shows the frequency variation of the
uniform mode as function of the magnetic field applied either along
$x$ direction (open symbols) either along $y$ direction (filled
symbols). The angular dependence of the resonance field of the uniform
mode at 10 GHz is also presented in Figure 2b. A non negligible uniaxial
anisotropy of magnitude $\sim310$ Oe is observed along $x$ direction
(\textit{i. e.} $\varphi_{H}=0$) corresponding to a minimum resonance
field. Since the studied Ni film is polycrystalline with no preferred
crystallographic orientations, no macroscopic magnetocrystalline effect
is expected, thus this uniaxial anisotropy should have a magnetoelastic
origin. Indeed, as the film is deposited on a flexible substrate,
a slight curvature along a given direction during the elaboration
process should lead to a non zero magnetoelastic anisotropy of the
maintained flat sample during FMR measurements, as suggested by Zhang\textit{
et al}. \cite{Zhang2013}. In this work, this uniaxial anisotropy
will be fitted by using an \textit{ad hoc} uniaxial anisotropy characterized
by an anisotropy constant $K_{U}$. From now, easy axis will refer
to $x$ direction since the angle between $x$ and the uniaxial anisotropy
is close to zero. Thus, the lines (continuous and dashed) in Figure
2 correspond to fits using the following expression obtained with
the assumption of an in-plane magnetization (\textit{i. e.} $\theta=\frac{\pi}{2}$):

\begin{figure}
\includegraphics[bb=25bp 320bp 605bp 555bp,clip,width=8.5cm]{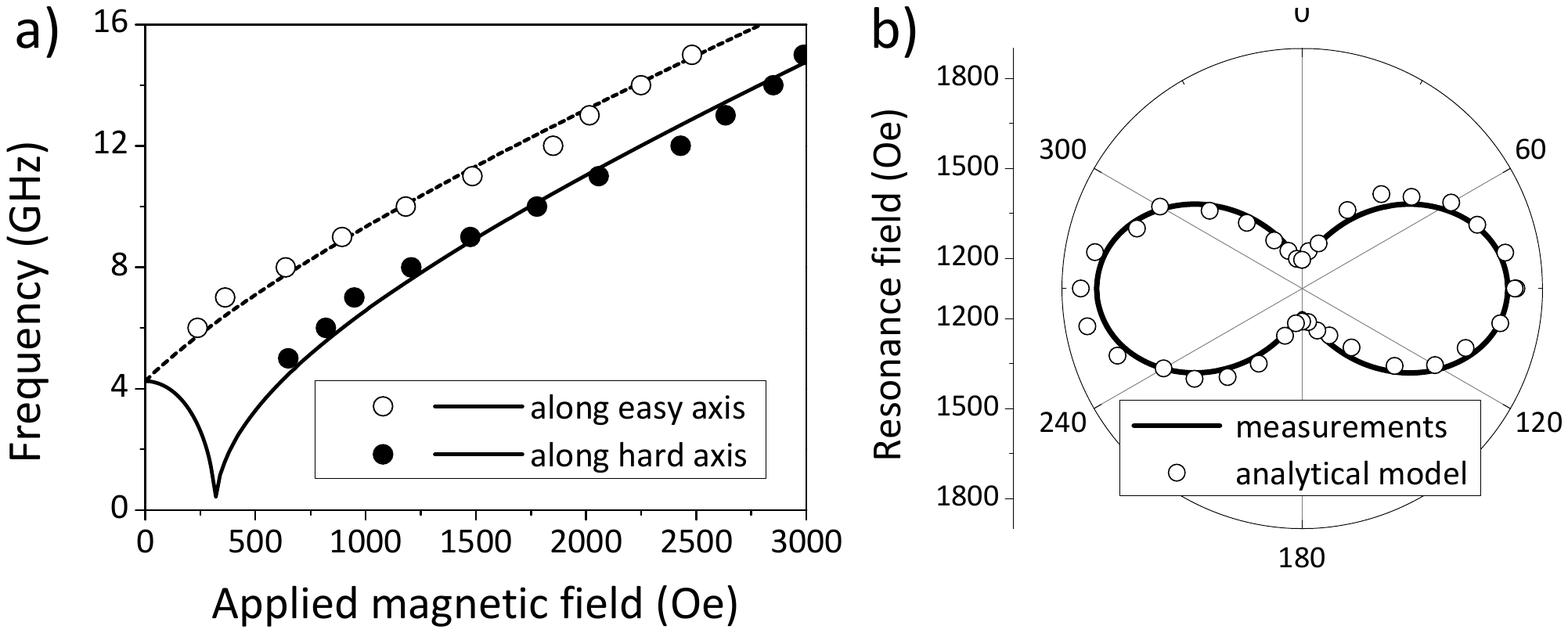}

\caption{a) Frequency variation of the uniform mode as function of the in-plane
applied magnetic field at zero applied voltage. Open symbols are obtained
with $\varphi_{H}=0$ ($\vec{H}\parallel\hat{x}$) and the filled
ones are obtained with $\varphi_{H}=\pi/2$ ($\vec{H}\parallel\hat{y}$).
b) In-plane angular dependence of the resonance field at 10 GHz. The
solid and dashed lines in a) and b) are best fits to the experimental
data using the following parameters: $\gamma=1.885\times10^{7}$ Hz.Oe$^{-1}$,
$M_{S}=480$ emu.cm$^{-3}$, $K_{U}=7.3\times10^{4}$ erg.cm$^{-3}$
and equation \ref{eq:frequency_0V}. }
\end{figure}
\begin{multline}
f^{2}=\left(\frac{\gamma}{2\pi}\right)^{2}\left(H_{res}\cos\left(\varphi-\varphi_{H}\right)+\frac{2K_{U}}{M_{S}}\cos2\varphi\right)\\
\left(H_{res}\cos\left(\varphi-\varphi_{H}\right)+\frac{2K_{U}}{M_{S}}\cos^{2}\varphi+4\pi M_{S}\right)\label{eq:frequency_0V}
\end{multline}

Where $f$ is the microwave driven frequency. It results from the
analysis of the data that the magnetic parameters are close to the
one of bulk Ni \cite{Neugebauer1959}. A magnetization at saturation
($M_{S}$) of $480$ emu.cm$^{-3}$ ($4\pi M_{S}\simeq6030$ G) with
a gyromagnetic factor ($\gamma$) of $1.885\times10^{7}$ Hz.Oe$^{-1}$
are found. The uniaxial anisotropy constant is estimated to $K_{U}=7.3\times10^{4}$
erg.cm$^{-3}$ (we define $H_{U}=2K_{u}/M_{S}\simeq315$ Oe). These
measurements performed at zero voltage will serve as reference in
order to evaluate the changes related to the in-plane strains induced
by the piezoelectric actuator.

\begin{figure}
\includegraphics[bb=66bp 225bp 545bp 588bp,clip,width=8.5cm]{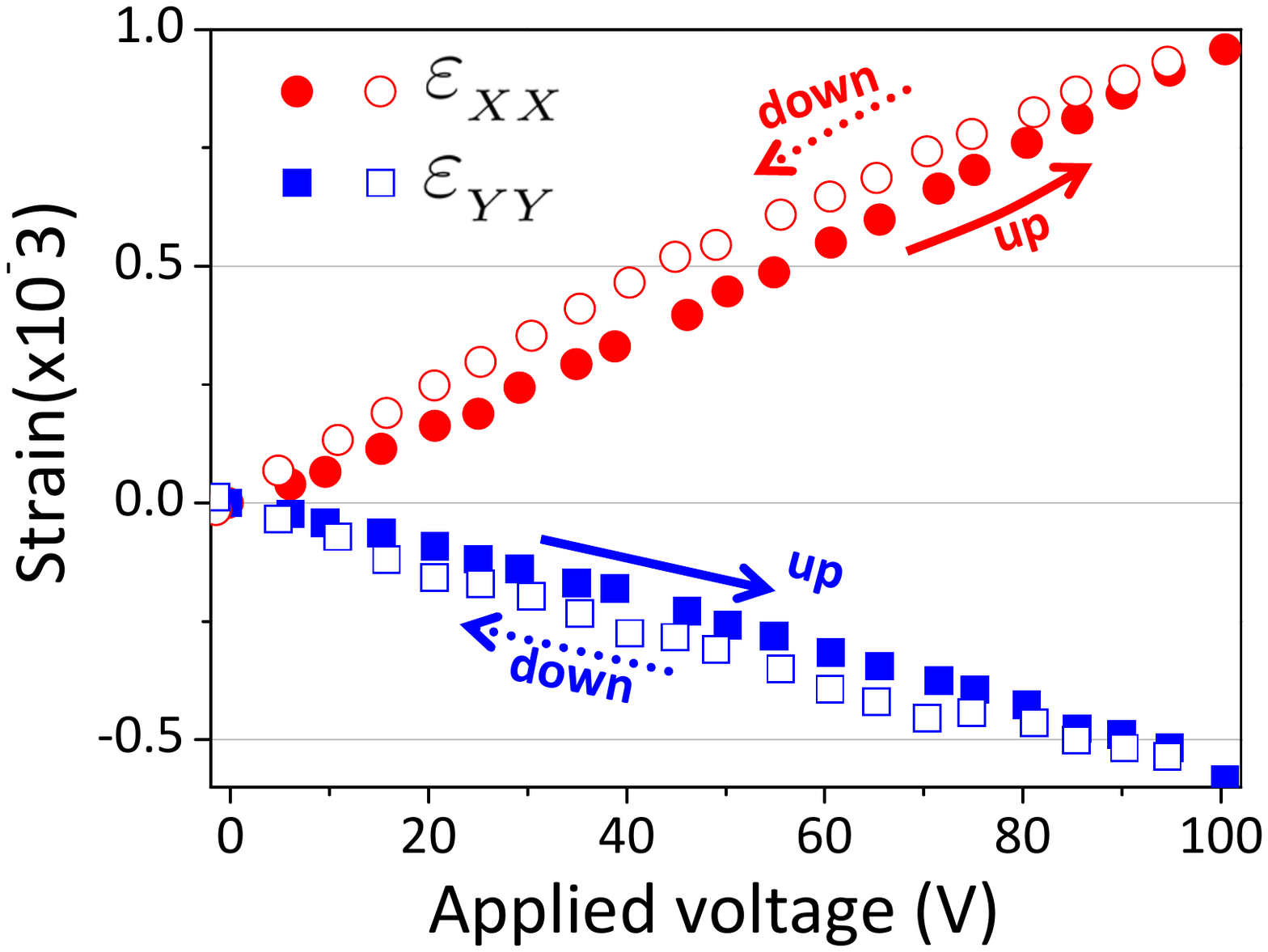}

\caption{DIC measurements of the mean in-plane strains ($\varepsilon_{XX}$
and $\varepsilon_{YY}$) at the surface of the Ni film. The circles
correspond to $\varepsilon_{XX}$ while squares correspond to $\varepsilon_{YY}$.
The applied external voltage is ranging from 0 V to 100 V (filled
symbols) and back to 0 V (open symbols) with steps of $\sim$5 V.
The first measurement at 0 V has been performed after ``saturating''
the actuator at 100 V to avoid training effect of the polarization.}
\end{figure}

\subsection{Quantitative in-plane strains measurements by DIC}

DIC technique has been used to measure the in-plane strains ($\varepsilon_{XX}$
and $\varepsilon_{YY}$) at the surface of the film. The measurements
have been performed by varying the external voltage from 0 V to 100
V (filled symbols) and back to 0 V (open symbols) with steps of around
5 V. Note that the first 0 V image was taken after applying a ``saturating''
voltage of 100V in order to avoid hysteresis effects which can occur
after several hours at zero voltage. The measured $\varepsilon_{XX}$
and $\varepsilon_{YY}$ as function of the applied voltage are shown
in Figure 3 and reach respectively $10^{-3}$ and $-0.5\times10^{-3}$
at 100 V. Indeed $\varepsilon_{XX}$ is found to be positive while
$\varepsilon_{YY}$ is found to be negative with a ratio $\left|\frac{\varepsilon_{YY}}{\varepsilon_{XX}}\right|\simeq0.5$.
Non linear variations for both $\varepsilon_{XX}$ and $\varepsilon_{YY}$
with the voltage is observed due to the intrinsic properties of the
ferroelectric material used in the fabrication of the actuator \cite{Pi_actuator}.
This behavior is well reproducible after ``saturating'' the actuator
a first time (at 100 V in this work). It should be noted that these
values correspond to mean values calculated in area close to $4\times4$
mm$^{2}$ (area of the thin film). However, the heterogeneties of
$\varepsilon_{XX}$ and $\varepsilon_{YY}$ can be neglected compared
to their ``absolute'' values and even at the edge of the $4\times4$
mm$^{2}$ area. Consequently, the measured $\varepsilon_{XX}$ and
$\varepsilon_{YY}$ can be considered as homogeneous in $XY$ plane.
Morever, the estimated in-plane shear strains by DIC were found to
be negligible.

In addition, the analysis of $\varepsilon_{XX}$ and $\varepsilon_{YY}$
on an uncoated area of the actuator gives same values. Thus, the transmission
of the in-plane strains between the actuator and the surface of the
film is close to $100\%$. Therefore, the values of $\varepsilon_{XX}$
and $\varepsilon_{YY}$ shown in Figure 3 should be homogeneous in
the thickness, \textit{i.e.} the strains are well transmitted through
the cement and through the film/substrate interface. This result is
consistent with previous studies performed on metallic films of comparable
thicknesses deposited onto flexible substrates and analyzed both by
DIC and x-ray diffraction techniques \cite{Djaziri2011,Djaziri2013}.

\begin{figure}
\includegraphics[bb=30bp 430bp 370bp 595bp,clip,width=8.5cm]{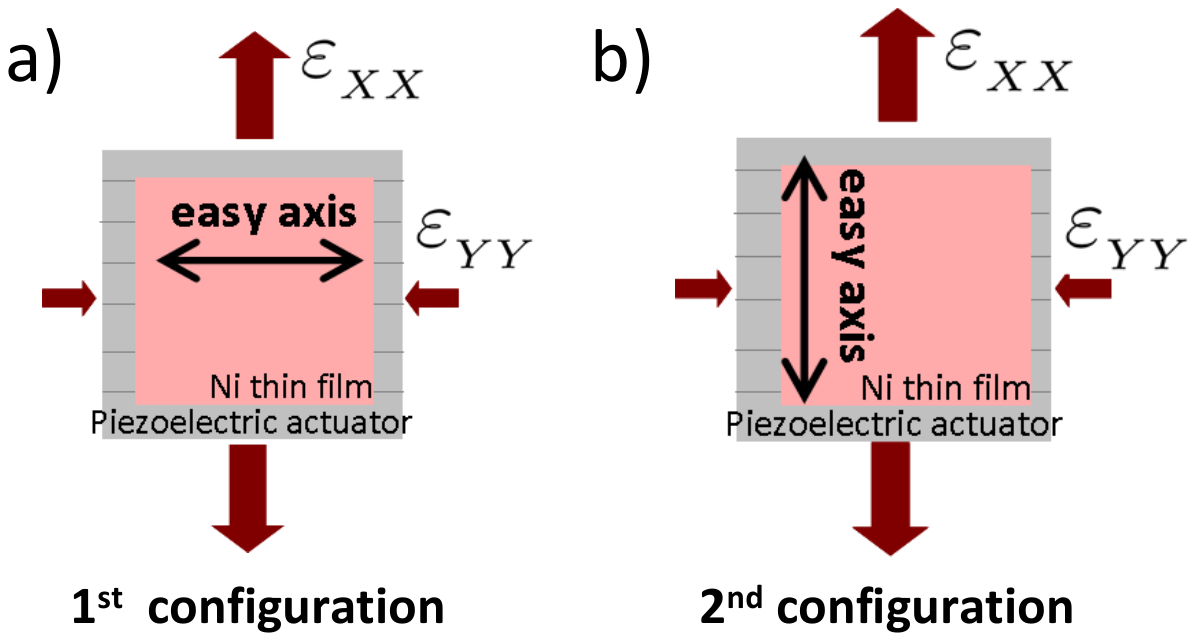}

\caption{a)b) Top view sketches of the two configurations used in the \textit{in
situ} MS-FMR investigations. }
\end{figure}

\begin{figure}
\includegraphics[bb=30bp 130bp 397bp 598bp,clip,width=8.5cm]{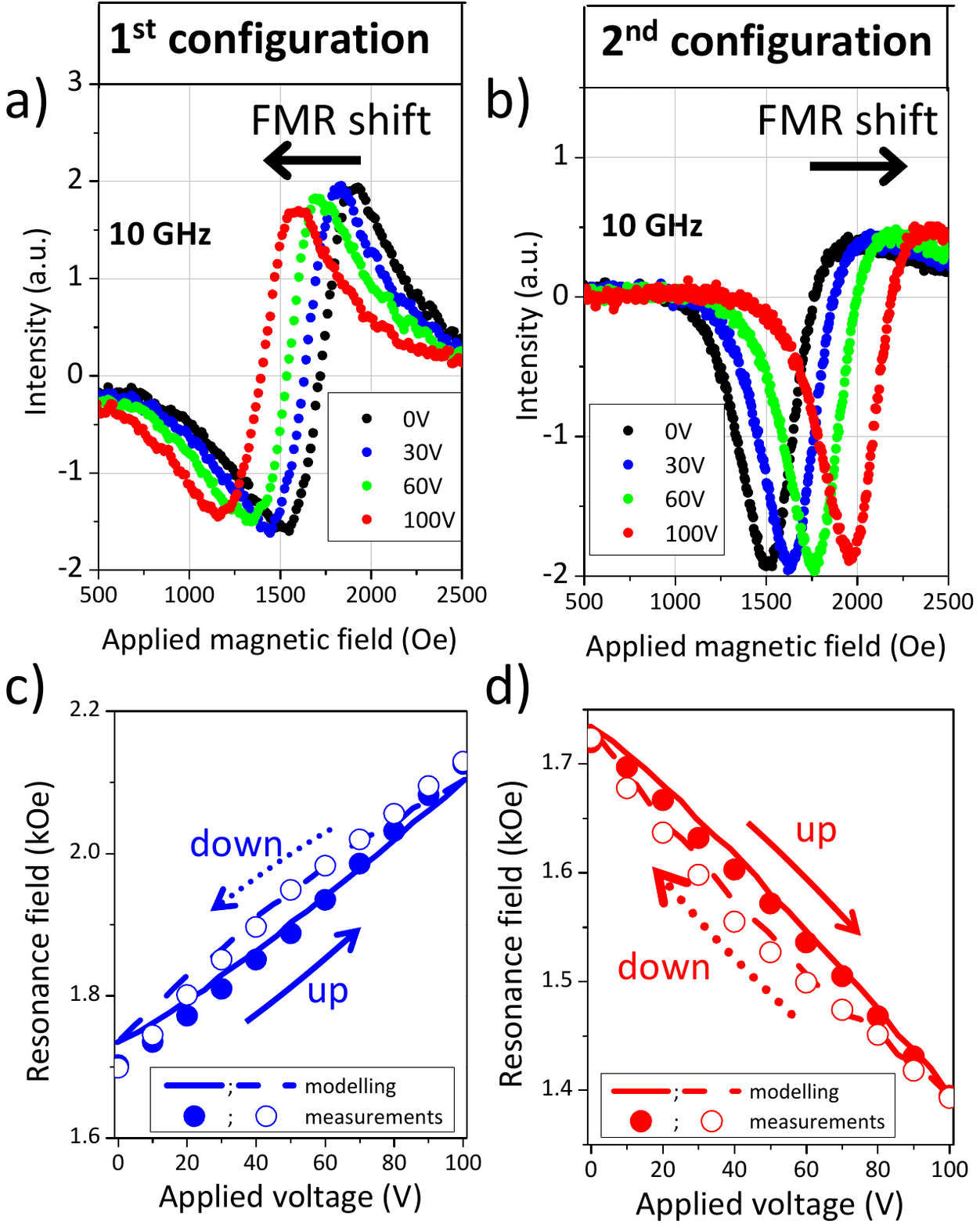}

\caption{a)b) Experimental spectra recorded at 10GHz with $\varphi_{H}=0$
at different voltages for the first and the second configurations
illustrated in Figure 4, respectively. c)d) Resonance field variations
(at 10 GHz) as function of the applied voltage with $\varphi_{H}=0$
for the first and second configurations, respectively. The open symbols
correspond to upsweep (0 to 100 V) while filled symbols correspond
to downsweep (100 to 0 V). The solid and dashed lines are best fits
to the experimental data by using an analytical expression deduced
from equations 2-4. An isotropic magnetostriction coefficient of $-26\times10^{-6}$
is deduced. }
\end{figure}

\subsection{Voltage induced magnetic anisotropy probed by MS-FMR}

The voltage induced anisotropy has been studied in two different configurations
defined by the angle between the easy axis and $X$ direction (see
Figure 4). In those configurations, the easy axis is either aligned
along $\hat{Y}$ (first configuration, Figure 4a) either along $\hat{X}$
(second configuration, Figure 4b). As previously mentioned, the Ni
film is polycrystalline with no preferred orientations. This allows
to define an effective isotropic magnetostrictive coefficient at saturation
$\lambda$ \cite{DuTr=0000E9mollet1993}. In these conditions and
because of the opposite sign of $\varepsilon_{XX}$ and $\varepsilon_{YY}$
values, an additional in-plane anisotropy field parallel to $Y$ direction
is expected (in the range 0-100V). Consequently, the initial easy
axis is expected to be increased in the first configuration while
a competition between the initial in-plane anisotropy and the voltage
induced one will takes place in the second configuration.

The influence of the applied voltage on the magnetic properties of
the thin film has been probed by MS-FMR technique. Figures 5a and
5b show typical MS-FMR experimental spectra recorded at 10 GHz with
an applied magnetic field along the initial hard axis (\textit{i.
e.} $\phi_{H}=0$ for the first configuration and $\phi_{H}=\frac{\pi}{2}$
for the second configuration) and for different applied voltages (0,
30, 60 and 100 V). Note that the experimental conditions (in term
of applied voltage) were same than for the DIC study. Moreover, in
order to avoid non linear variations of $\lambda$ as function of
the applied magnetic field \cite{DuTr=0000E9mollet1993}, the experimental
spectra have been recorded at 10 GHz. In these conditions, the deduced
resonance fields are in a magnetic saturating regime (as shown in
Figure 2).

It clearly appears that the resonance field increases (resp. decreases)
with the applied voltage in the first (resp. second) configuration.
This indicates that the initial uniaxial anisotropy is either reinforced
in the first configuration either softened in the second one. This
is consistent with a negative magnetostriction coefficient \cite{Brandlmaier2009}.
The deduced resonance fields as function of the applied voltage are
presented in Figures 5c and 5d. Here, filled symbols represent the
upsweep (0 to 100 V) and open symbols the downsweep of the applied
voltage (100 to 0 V). The observed non-linear and hysteretic variations
of the resonance field is due to the variations of $\varepsilon_{XX}$
and $\varepsilon_{YY}$ as function of the applied voltage (see Figure
3).

In order to quantitatively bind the resonance field variations to
the in-plane strains induced by the applied voltage, we can calculate
the uniform precession mode frequency as function of the applied voltage
(strain) by adding a magnetoelastic energy term $F_{ME}$ to the total
magnetic energy density $F$ of the Ni film. In an isotropic case,
which is closely ours (non-textured Ni film), $F_{ME}$ energy term
can be written as:

\begin{equation}
F_{ME}=-\frac{3}{2}\lambda\left(\left(\gamma_{x}^{2}-\frac{1}{3}\right)\sigma_{xx}+\left(\gamma_{y}^{2}-\frac{1}{3}\right)\sigma_{yy}\right)
\end{equation}

$\sigma_{xx}$ and $\sigma_{yy}$ being the in-plane principal stress
tensor components while $\gamma_{x}$ and $\gamma_{y}$ correspond
to the direction cosines of the in-plane magnetization. Moreover,
the relation between the principal components of stress and strain
tensors is given by the following equations where $E$ is the Young's
modulus and $\nu$ is the Poisson's ratio:

\begin{equation}
\sigma_{xx}=\left(\frac{E}{1+\nu}\right)\left(\frac{1}{1-\nu}\varepsilon_{xx}+\frac{\nu}{1-\nu}\varepsilon_{yy}\right)
\end{equation}

\begin{equation}
\sigma_{yy}=\left(\frac{E}{1+\nu}\right)\left(\frac{1}{1-\nu}\varepsilon_{yy}+\frac{\nu}{1-\nu}\varepsilon_{xx}\right)
\end{equation}

\begin{figure}
\includegraphics[bb=46bp 190bp 545bp 595bp,clip,width=8.5cm]{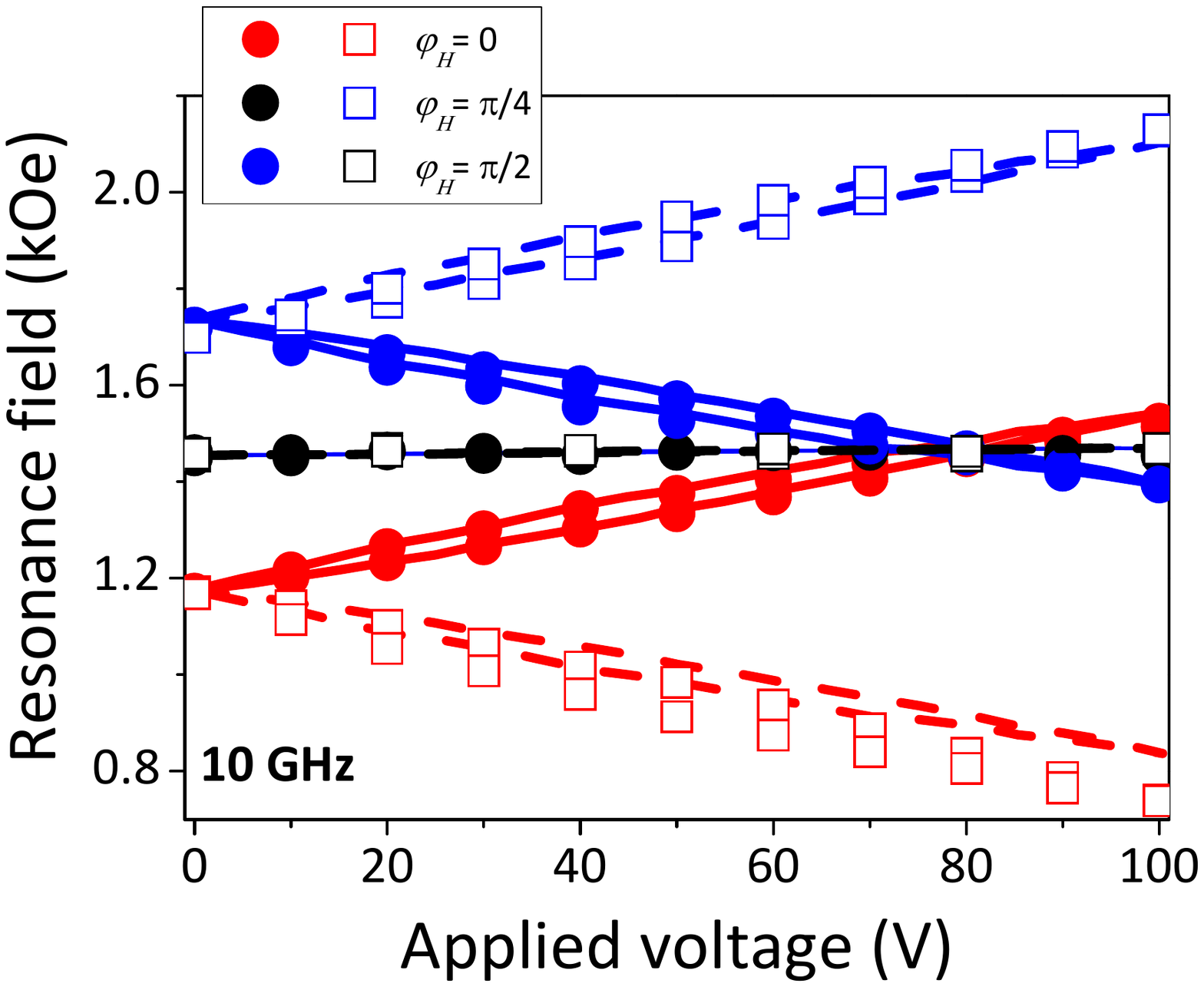}

\caption{Resonance field as function of the applied voltage at 10 GHz for different
$\varphi_{H}$ values ($0,\frac{\pi}{4}$ and $\frac{\pi}{2})$. Filled
symbols correspond to the experimental data obtained from the first
configuration (Figure 4a) while open symbols are obtained from the
second configuration (Figure 4b). The dashed and continuous lines
are fits to the experimental data obtained by using an analytical
expression deduced from equations 2-4. A magnetostrictive coefficient
at saturation of magnitude $\lambda=-26\times10^{-6}$ is deduced. }
\end{figure}

The voltage dependence of the uniform mode resonance field is calculated
by using the following relation \cite{Netzelmann1990}: $\left(\frac{2\pi f}{\gamma}\right)^{2}=\frac{1}{\left(M\sin\theta\right)^{2}}\left(\frac{\partial^{2}F}{\partial\theta^{2}}\frac{\partial^{2}F}{\partial\varphi^{2}}-(\frac{\partial^{2}F}{\partial\theta\partial\varphi})^{2}\right)$.
Assuming a Young's modulus of $2\times10^{10}$ dyn.cm$^{-2}$ ($\equiv200$
GPa) and a Poisson's ratio of 0.3 \cite{Smithells1976}, the only
undetermined parameter is $\lambda$. Introducing the experimentally
found voltage dependence of $\varepsilon_{XX}$ and $\varepsilon_{YY}$,
the best fits of all the experiments performed in the both studied
configurations gives $\lambda=-26\times10^{-6}$. It should be noted
that this value is consistent with commonly found values for non-textured
polycristalline Ni film (about $-30\times10^{-6}$) and validate our
overall approach. It should be noted here that the polycristalline
theoretical value can be found by applying an arithmetic average of
the analytical solutions given by Reuss and Voigt models \cite{Callen1965}
:

\begin{equation}
\lambda^{Reuss}=\frac{2}{5}\lambda_{100}+\frac{3}{5}\lambda_{111}
\end{equation}

\begin{equation}
\lambda^{Voigt}=\frac{2}{2+3A}\lambda_{100}+\frac{3A}{2+3A}\lambda_{111}
\end{equation}

Where $\lambda_{100}$ and $\lambda_{111}$ are respectively the magnetostriction
coefficients of $<100>$ and$<111>$ directions families (respectively
equal to $-51\times10^{-6}$and $-22.4\times10^{-6}$), $A$ is the
Zener anisotropy index ($A=2C_{44}/(C_{11}-C_{12})\cong2.6$ for Ni)
\cite{DuTr=0000E9mollet1993}.

Note that the hysteretic and non-linear variations of the resonance
field as function of the applied voltage is well reproduced by the
analytical model proposed.

Figure 6 shows variations of the resonance field as function of the
applied voltage for different $\varphi_{H}$ angles ($0,\frac{\pi}{4}$
and $\frac{\pi}{2}$). Here, open squares correspond to experimental
data obtained from the first configuration while filled circles are
obtained from the second configuration. The measurements coming from
the first configuration indicate that the voltage induced anisotropy
is parallel to the initial uniaxial one and consequently the magnitude
of the easy axis along $Y$ is increased. This behavior is observed
\textit{via} the difference between the resonance fields measured
at $\varphi_{H}=0$ and $\varphi_{H}=\frac{\pi}{2}$ which is roughly
$2H_{U}\sim600$ Oe at zero voltage and around 1400 Oe at 100 V. Thus,
the magnitude of the easy axis at 100V is equal to more than twice
time the measured value in the absence of voltage. The second configuration
is more interesting since a competition between the initial uniaxial
anisotropy and the voltage induced one takes place. It results that
the initial anisotropy is totally compensated at around 80 V (\textit{i.
e.} $H_{ME}=H_{U}$, where $H_{ME}$ being the induced magnetoelastic
anisotropy field) which leads to an isotropic in-plane behavior of
the resonance field. Interestingly, at 100V a non zero easy axis perpendicular
to the initial one at 0 V is observed.

The solid and dashed lines in Figure 6 are fits to the experimental
data calculated by using the parameters previously given. One can
note the good agreement between the experimental measurements and
the calculated lines.

Note that a non-negligible overestimation of the magnetostrictive
coefficient can occurs if we neglect the influence of $\varepsilon_{YY}$
(\textit{i. e.} asumption of a uniaxial strain). Indeed, the calculated
lines (not shown here) in this approximation are quasi similar to
the one previoulsy calculated. However, an overestimation of $\lambda$
is performed ($\lambda=-40\times10^{-6}$ instead of $-26\times10^{-6}$).
Thus, the present study also demonstrates that a full characterization
of the in-plane strains is needed if we want to quantitatively analyzed
the inverse magnetoelectric effect occuring inside such structures.

\section{Conclusion}
\begin{itemize}
\item Magnetic anisotropy in Nickel thin films grown on Kapton\textregistered{}
substrate has been studied by microstrip ferromagnetic resonance.
An ``initial'' uniaxial ansiotropy has been found and is most probably
related to the non-flatness of the compliant substrate during deposition.
\item Piezo-actuation of Ni/polyimide system has been performed allowing
a voltage-induced strain control leading to a change of the total
magnetic anisotropy.
\item An accurate study by digital image corelation technique allowed a
good knowledge of in-plane macroscopic strains at the surface of the
film as function of applied voltage to the piezoelectric actuator.
This technique that allows high precision (about $5\times10^{-5}$)
is found to be relevant for determining in-plane strains in the elastic
domain.
\item A comparison of strains of the film with that of the piezoelectric
actuator demonstrates the full transmission of in-plane elastic strains
across the whole system, which is due to the compliant behavior of
the polyimide substrate.
\item A simple analytical model taking into acount of the measured strains
as function of applied voltage well matches the experimental data.
The only adjusted parameter is the effective magnetostrictive coefficient
at magnetic saturation of Ni film and is found to be close to previously
published values.\end{itemize}
\begin{acknowledgments}
The authors gratefully acknowledge the French Research Agency (ANR)
for their financial support (project ANR 2010 JCJC 090601 entitled
``SpinStress''). The authors also acknowledge fi{}nancial support
of the CNRS in the framework of the ``PEPS INSIS'' program (FERROFLEX
project). The authors thank Tarik Sadat (PhD student at Paris 13th
university) for sharing his automatic data analysis program.\end{acknowledgments}

\end{document}